\documentstyle[12pt]{article}
\begin{document}

\begin{center}

{\Large \bf Low Frequency Quantum Transport in a Three-probe 
Mesoscopic Conductor}

\bigskip

Qingrong Zheng and Jian Wang 

\bigskip

{\it
Department of Physics, \\
The University of Hong Kong,\\
Pokfulam Road, Hong Kong.
}

\bigskip

Hong Guo

\bigskip

{\it
Centre for the Physics of Materials,\\
Department of Physics, McGill University,\\
Montreal, Quebec, Canada H3A 2T8.
}

\end{center}

\vfill

\baselineskip 15pt               

The low frequency quantum transport properties of a three-probe mesoscopic 
conductor are studied using B\"uttiker's AC transport formalism.
The static transmission coefficients and emittance matrix of the system 
were computed by explicitly evaluating the various partial density of 
states (PDOS). We have investigated the finite size
effect of the scattering volume on the global PDOS. By increasing the
scattering volume we observed a gradual improvement in the agreement of 
the total DOS as computed externally or locally. Our numerical data permits
a particular fitting form of the finite size effect.

\vfill

\baselineskip 16pt


\newpage
\section{Introduction}

In recent years extensive investigations of ballistic and mesoscopic
quantum conductors have been carried out\cite{review}.  Experimentally
the advances in nanotechnology have enabled the possibility of fabricating
submicron structures with linear size of 1000 Angstrom or less. 
Due to quantum size effect, the transport properties of these small systems
can be very different from their classical counterpart and many interesting
phenomena have been discovered\cite{review}. On the theoretical side,
a main tool for understanding {\it static} ballistic transport is based 
on the scattering approach of Landauer-B\"uttiker 
formalism\cite{landauer,but1}. 

It has been realized that the usual static scattering approach 
can not be directly applied to {\it dynamic} transport problems where 
the external potential has a time dependent oscillating component. As shown by 
B\"uttiker and his co-workers, a direct application of the original approach 
of Landauer-B\"uttiker formalism can not yield electric current and charge
conservation. To preserve this conservation, it is necessary to consider 
the implication of the long range Coulomb interaction.  As a result the
AC transport theory for coherent quantum conductors are more complicated.
At present there are several approaches to deal with the problem of 
computing AC conductance. In a strongly correlated electronic system, 
the Anderson impurity model is often used. To treat AC transport one 
employs the linear response theory in conjunction with the Keldysh 
Green's function which is often applied to deal with non-equilibrium 
problems\cite{keldysh}. One can also use Kubo's linear response theory 
by assuming that the electric field inside the sample is known 
{\it a priori}. However this is a very strong requirement\cite{avishai}. 

Along another line of development B\"uttiker and his co-workers have 
advanced a current conserving formalism\cite{but2}. The 
key idea in this theory is to consider the self-consistent internal potential
so that the current and the charge is conserved. In a series of articles, 
B\"uttiker and co-workers investigated several low frequency quantum 
transport problems\cite{but2,but3,but4,but5,but96,butp1}. To first order in 
frequency $\omega$, the response of an arbitrary scattering problem to
quasi-static perturbations in the scattering potential is naturally 
expressed in terms of a set of local partial density of states (PDOS)
each associated with one element of the scattering matrix. This AC transport 
formalism has also been extended to second order in frequency $\omega$ in 
the quantum hall regime\cite{butp1}.

The application of B\"uttiker's AC transport theory is easier in 1D,
such as a 1D quantum well, a $\delta$-function potential, and a perfect 
quantum wire, where the scattering matrix and wave function can be 
obtained analytically. Much intuition and interesting results have been
obtained from 1D calculations which can often been done analytically. 
The investigations for 2D conductors have recently 
begun\cite{wang1,wang2,wang3,wang4} and the dynamic transport response of a
3D atomic wire has been calculated using first principles\cite{wan}.
However, due to technical difficulties, in 2D one usually can not obtain 
analytical expressions for the AC admittance except for very special 
cases\cite{wang2,wang3}. For more general situations numerical calculations 
are needed.

It is the purpose of this article to further investigate AC admittance of
2D coherent conductors in the ballistic regime. In particular we shall
focus on numerically analyzing a three probe conductor as shown in Figure
(1). There are several motivations for this study. First, 
similar to that of the DC transport situation\cite{review}, we believe 
coherent AC admittance of multi-probe (by multi-probe we mean more than 
two probes) conductors should be studied in detail since usually 
experimental measurements are conducted in multi-probe setups. 
However to the best 
of our knowledge there is as yet a detailed numerical analysis of any 
2D multi-probe systems. Second, in our investigations of two-probe 
conductors\cite{wang1,wang2}, an important technical point is the size 
effect of the scattering volume. It was found\cite{gas1,wang1,wang2} 
that the total PDOS as computed from {\it external} global PDOS 
(GPDOS) does not equal to that computed from the {\it local} PDOS 
(LPDOS), unless the scattering volume is very large. This led to
a violation of the current conservation and gauge invariance in numerical
calculations where the scattering volume is always finite. Hence there is a
need to explicitly and systematically examine the scattering volume size 
effect. Finally, in order to study certain physical effects such as the
inelastic and dissipative effects using the quantum scattering approach,
a very useful phenomenological approach is to introduce fictitious links
from the conductor to external dissipative reservoirs\cite{but0}.  In this
case one must deal with multi-probe situations.

To these purposes we have computed the low frequency admittance of a
three-probe junction (see Fig. (1)). We have examined the 
emittance matrix for both the tunneling regime and the transmissive 
regime in detail. The behavior of the emittance matrix is found
to be closely related to that of the transmission coefficients. 
We have computed the total PDOS from both the {\it external} GPDOS and 
the {\it local} PDOS, and largely speaking the two total PDOS as obtained 
approach each other as the scattering volume is increased. We found that there
exists a ``critical region'' in energy near the second propagating subband
threshold, such that within this region the charge conservation is not
strictly obeyed for any finite scattering volume. However the larger the
scattering volume, the smaller the ``critical region'' is.

This paper is organized as the following. In the next section we set out
the theoretical and numerical procedures of computing the transmission 
functions. In section III we present and analyze the numerical results. 
Both the dynamic and static transport properties and their relationship 
will be discussed. Finally, a summary is given in section IV. 

\section{Theoretical and numerical analysis}

The current conserving dynamic transport formalism proposed by B\"uttiker and
co-workers is amply reviewed in several articles\cite{buttiker1a} and we
refer interested readers to them. In this section we shall only outline
our theoretical and numerical procedures to compute the necessary quantities 
such as the various partial density of states for the 3-probe system.  

It has been shown by B\"uttiker, Thomas, Pr\^etre, Gasparian and 
Christen\cite{but1,but2,but3,gas1}, to linear order in frequency $\omega$
the admittance is given by the following formula,
\begin{equation}
g_{\alpha \beta }^I(\omega )\ =\ g_{\alpha \beta }^e(\omega
=0)\ -\ i\omega e^2 E_{\alpha\beta}
\label{gfinal}
\end{equation}
where the emittance matrix $E_{\alpha\beta}$ is calculated from the various
partial density of states:
\begin{equation}
E_{\alpha\beta}\ =\ 
\left( \frac{dN_{\alpha \beta }}{dE}-D_{\alpha\beta}\right)\ \ \ .
\label{em1}
\end{equation}
The subscripts $\alpha\beta$ indicates scattering from a lead labeled by 
$\beta$ to that labeled by $\alpha$. The first term in the emittance matrix
gives the AC response of the system to the external potential change, while
the second term is from the internal potential change induced by the
external perturbations.  The external contribution is determined by the
global partial density of states\cite{but5}: for a large scattering volume
the global PDOS can be expressed in terms of the energy derivative of the 
scattering matrix elements\cite{band}
\begin{equation}
\frac{dN_{\alpha \beta }}{dE}\ =\ \frac 1{4\pi i}\left(
s_{\alpha \beta }^{\dagger }\frac{ds_{\alpha \beta }}{dE}\ -\ \frac{%
ds_{\alpha \beta }^{\dagger }}{dE}s_{\alpha \beta }\right)\ \ \ .
\label{gpdos}
\end{equation}
On the other hand the internal contribution $D_{\alpha\beta}$ is related to
the local PDOS, and within the Thomas-Fermi linear screening model is given
by
\begin{equation}
D_{\alpha \beta}\ \equiv \ \int d{\bf r}\left[ \frac{dn(\alpha ,%
{\bf r})}{dE}\right] \left[ \frac{dn({\bf r})}{dE}\right] ^{-1}\left[ \frac{%
dn({\bf r},\beta )}{dE}\right]\ \ ,
\label{dab}
\end{equation}
where the local PDOS (called {\it injectivity}) is calculable from the 
scattering wavefunction,
\begin{equation}
\frac{dn({\bf r},\alpha )}{dE}\ =\ \frac 1{hJ}|\Psi_{\alpha} ({\bf %
r})|^2\ \ .
\label{inject}
\end{equation}
where $J$ is the incident flux and $\Psi_{\alpha}(\bf {r})$ is the
scattering wavefunction for electrons coming from the probe $\alpha$. 
In the absence of a magnetic field, the {\it emissivity} $dn(\beta,{\bf r})/dE$
equals to the injectivity\cite{butp1}. Finally, $dn({\bf r})/dE = 
\sum_{\alpha} dn(\alpha,{\bf r})/dE$ is the total local density of states.
It is straightforward to prove that the current is conserved since the
admittance matrix $g^I$ satisfies 
$\sum_\alpha g_{\alpha \beta }^I(\omega )=0$. 
This can be seen by realizing that $\sum_\alpha dN_{\alpha \beta
}/dE\equiv d\bar N_\beta /dE$ is the injectance which is identical to 
$\sum_\alpha D_{\alpha \beta }$.

To compute the various PDOS, for simplicity we shall focus on the first 
{\it transport} subband only, thus the incoming electron energy is 
restricted to within the interval $(\pi/a)^2<E<(2 \pi/a)^2$ in units 
of $\hbar^2/(2ma^2)$ with $m$ the effective mass of the electron and 
$a$ the width of the leads (see Fig. (1)).  Multi-subbands can also be
included without difficulties, such as that of Ref. \cite{wan}.
The scattering properties of the three-probe system is then
characterized by a $3 \times 3$ scattering matrix 
${\bf S}(E)\equiv\{s_{\alpha \beta}\}$ with $\alpha$, $\beta=1,2,3$.
For example, for an incident electron coming from probe 1, it scatters in the
scattering volume, and then reflects back to probe 1 with a probability 
amplitude given by $|s_{11}|$, or transmits to probes 2 and 3 with 
probability amplitudes $|s_{21}|$ and $|s_{31}|$, respectively. 
The transmission coefficients can thus be expressed in terms of 
scattering matrix, {\it i.e.} $T_{\alpha \beta} = |s_{\alpha \beta}|^2$.
For the system of figure (1), the scattering matrix has the following 
symmetry: $|s_{11}|=|s_{22}|,$ $|s_{21}|=|s_{12}|,$ $|s_{31}|=|s_{32}|,$
$|s_{13}|=|s_{23}|$ and $|s_{13}|=|s_{31}|$. Therefore, there are only four 
distinct elements out of nine.

For the three-probe conductor of Figure (1), the
quantum scattering problem is solved using a mode matching method.  
The wavefunction in region I can be written as
\begin{equation}
\Psi_I = \sum_n \chi_n(y) (a_n e^{i k_n x} +b_n e^{-i k_n x}) ,
\label{psi1}
\end{equation}
where $\chi_n(y)$ is the transverse wave function, $k_n^2 = E-(n\pi/a)^2$ is 
the transport energy, $a_n$ is the input parameter, and $b_n$ is the 
reflection amplitude. Similarly for region II, we have
\begin{equation}
\Psi_{II} = \sum_n \chi_n(y) (c_n e^{i k_n x} +d_n e^{-i k_n x}) \ \ .
\label{psi2}
\end{equation}
For region III,
\begin{equation}
\Psi_{III} = \sum_n \chi_n(x) (e_n e^{i k_n y} +f_n e^{-i k_n y}) ,
\label{psi3}
\end{equation}
where $c_n$ and $e_n$ are transmission amplitudes and $d_n$ and $f_n$
are input parameters. 
The wavefunction in region IV is the combination of wavefunctions in
regions I, II, and III. 
At the boundaries of the various regions, we match the
wavefunctions and their derivatives and this leads to the desired 
transmission coefficients with which the scattering wave functions
Eqs. (\ref{psi1})-(\ref{psi3}) are also determined. 

If we choose point $O$ as the origin (see figure 1), the scattering matrix 
$s_{1\beta}$ is defined as
\begin{eqnarray}
s_{11} & = & b_1 \nonumber \\
s_{12} & = & c_1 e^{ik_1 a} \nonumber \\
s_{13} & = & e_1 e^{ik_1 a}  \ \ .
\end{eqnarray}
Using Eqs. (\ref{em1})-(\ref{inject}) and the solution of the scattering
problem, we can explicitly compute the low frequency admittance.

\section{Results}

We have investigated the transmission coefficient and the emittance matrix 
in two different transport regimes for various system parameters.   
The first regime is very transmissive and the second is a tunneling 
regime where tunnel barriers are added at the probes. 
The AC response of these regimes 
can be quite different as a transmissive situation tends to be
inductive, while a non-transmissive case tends to be capacitive 
(see below). The low frequency admittance is given by 
Eq. (\ref{gfinal}) in which the DC conductance $g_{\alpha\beta}^e(\omega=0)$ 
of our three-probe system is determined using transmission coefficients by 
applying the B\"uttiker multi-probe conductance formula\cite{but1}.

\subsection{Emittance}

In Figure (2) we show the transmission coefficients and the emittance 
$E_{\alpha\beta}$ in the transmissive regime as a function of the incoming
electron energy.  In this case the system does not show any resonance 
behavior and the transmission coefficients $T_{\alpha\beta}(E)$ are quite
large for most of the energy range while the reflection coefficient $R_{11}$
is small (Fig. (2a)).  It is interesting to find that the {\it shape} of
emittance are similar to that of the corresponding transmission coefficients,
as shown in Fig. (2b).  This is different from cases where quantum
resonances dominant the transport\cite{wang1} (see below) and for that case
the AC responses follow the DC transmissions only at the resonances.  

There are two different responses to the external time varying potential: 
capacitive and inductive depending on the sign of the emittance matrix 
element $E_{11}$. According to Eq.(\ref{em1}), $E_{11}$ consists of two 
terms: $dN_{11}/dE$ the capacitive term and $D_{11}$ the inductive term. 
For a two probe {\it capacitor} there is no DC current so that 
$dN_{12}/dE =0$. As a result $E_{12}$ is negative. Therefore for a capacitor 
$E_{11} = -E_{12}$ is positive. Extending this notion, one concludes that 
the system responds capacitively if $E_{11}$ is positive. For a ballistic 
conductor with complete transmission $dN_{11}/dE$ vanishes and $E_{11}$ 
is negative. In other words, negative $E_{11}$ gives an inductive response. 
These different responses are clearly shown in Fig. (2).

The AC transport properties are very different in the tunneling regime.
To establish such a regime, we have put tunneling barriers inside probes 
1 and 2 at the junctions between the probes and the scattering volume.
In particular the barrier heights are $V_{barrier}=40E_1$, and the width is
$0.1$ where the width of the wire $a$ has been set to one.  No barrier is 
added in probe 3.  We have also included a potential 
well with depth $V_{well}=-40E_1$ in the center of the scattering volume 
with a size of $2.8 \times 1.9 $.  The well and barriers establish several
transport resonances, these are clearly marked by the sharp peaks in the
electron dwell time defined as\cite{but7}
\begin{equation}
\tau_{\alpha} = \frac{1}{J} \int_{\Omega} |\Psi_{\alpha}({\bf r})|^2 
d^3 {\bf r} \ \ ,
\end{equation}
where $\Omega$ is the scattering volume. $\tau_1$ is plotted against energy
in Fig. (3) while the inset shows $\tau_3$. The dwell time measures the 
duration an electron spends in the scattering volume. Thus if transport 
is mediated by resonance states we expect much longer dwell 
times\cite{wang5} at the resonances. This idea has recently been proved 
by Iannaconne\cite{ian}.  Fig. (3) shows that three resonance states,
with energies $E_1=13.2$, $E_2=24.1$ and $E_3=35.6$ are established.
The quantum resonances also leads to sharp 
peaks in the transmission coefficient $T_{21}$ and reflection coefficient
$R_{11}$, as shown by the solid lines of Fig. (4a,4b). 
At these resonances both the GPDOS and LPDOS take maximum values, 
leading to the sharp jumps in the emittance $E_{11}$ and
$E_{21}$ as shown by the data points in Fig. (4a,4b). 
The variations of $E_{11}$ and $E_{21}$ as 
functions of energy $E$ are very closely correlated with those of
$R_{11}(E)$ and $T_{21}(E)$ near the resonances, as shown by Fig. (4).
Since there was no tunneling barrier in probe 3, the resonance 
transmission to that probe is not as sharp, and the transport behavior 
shows a mixture of tunneling and transmissive nature, as shown in Fig. (4c).

In the tunneling regime the AC response changes sharply from inductive 
at one side of the resonance energy to capacitive on the other side of 
the resonance or vise versa, in distinctive difference as compared to 
the transmissive case discussed above. Let's examine $E_{11}$ 
near resonance $E_3$. As the energy approaches to $E_3$, the system 
first responds inductively and is followed by a strong capacitive 
response. This behavior is clearly related to the fact that
the resonance is characterized by a complete {\it reflection} indicated by
the large peak in the reflection coefficient (see Fig. (4a)).
This behavior has been seen previously in 2D quantum wires\cite{wang2}.
On the other hand, for 1D resonance tunneling, a Breit-Wigner type 
{\it transmission} resonance gives rise to the similar AC response 
behavior\cite{but3} discussed here. When the incident energy is 
near the resonance $E_1$, the AC response is reversed: 
first capacitively and then inductively.  Hence the behavior near 
$E_1$ and $E_3$ are very different. For energy near $E_1$ 
the emittance behaves like an odd function but near $E_3$ it is like an 
even function. The reason, as we have checked numerically, is that the 
external and the internal responses do (not) reach the maximum at the 
same energy for $E$ near $E_3$ ($E_1$). This behavior of 
$E_{11}$ is also a manifestation of the reflection coefficient $R_{11}$.
As energy sweeps through $E_1$, the strong capacitive AC response is due to 
the complete reflection peak, and the following inductive response 
is because the reflection coefficient $R_{11}\approx 0$. 
Hence in the AC response of a system, near a quantum resonance
whether it is voltage following current (capacitive) first, or current 
following voltage (inductive) first, can only be determined by detailed
analysis and the outcome depends on the peculiarities of the system such
as the existence of a third probe as we have studied here.
In Figure (4c) we show the emittance matrix elements $E_{13}$. Although
they have much smaller values they do exhibit dips around three resonant 
energies $E_1$, $E_2$, and $E_3$.

\subsection{Finite size effect of the GPDOS}

A very important formal advance of the AC transport theory is the correct
characterization of electric current conservation.  In principle this
requirement is satisfied by the AC transport formalism used 
here\cite{but2} which demands $\sum_{\alpha}E_{\alpha\beta}=0$.  Hence
we must have 
$\sum_{\alpha}dN_{\alpha\beta}/dE=\sum_{\alpha}D_{\alpha\beta}$.
Since both sides of this equation represent the total scattering
DOS, thus the current conservation is obtained.   
In practical calculations, the left hand side of this equation is computed
{\it externally}, using the scattering matrix which is calculated at the
boundaries of the scattering volume. On the other hand the right side
of this equation is calculated {\it locally}, using the scattering 
wavefunction inside the scattering volume. These two quantities becomes
precisely equal when the scattering volume is very large\cite{buttiker1a}.
For a finite scattering volume, correction terms should be added to the
GPDOS, as shown in Ref. \cite{gas2} for 1D systems, and in Ref. \cite{wang2} 
for a 2D system.  Without the corrections, numerical results for a 2D 
quantum wire showed a small but systematic deviation from the precise 
current conservation\cite{wang1}.  Such a deviation actually diverges
at the edges of successive propagation subbands as shown in Ref. 
\cite{wang2}.

Since partial density of states play a vital role in the AC transport
formalism used in this work, in this section we present a detailed analysis
of the finite size effect of the scattering volume to the GPDOS. 
To this purpose we have examined a variety of system sizes $L$ for many
energies near the onset of the second transport subband.  
As a measure, we define a quantity which is the difference of the total 
DOS as calculated from GPDOS and LPDOS:
\begin{equation}
\delta_\beta\ \equiv\ \sum_{\alpha}\left(
\frac{dN_{\alpha\beta}}{dE}\ -\ D_{\alpha\beta}\right)\ \ \ .
\label{delta}
\end{equation}
Obviously $\delta_{\beta}=0$ if the current is precisely conserved.

Fig. (5a) shows $\delta_1$ as a function of the system size $L$ for three 
energies very close to the second subband edge which is located at 
${\cal E}_2=39.4784$. A clear crossover to the large volume limit is
revealed as $\delta_1\rightarrow 0$ when $L$ is increased. It is also
clear that for energy closer to ${\cal E}_2$, the crossover is slower 
(solid line).  We found that the decay of $\delta_1$ is essentially
exponential for all energies examined, and has a interesting form
for large $L$:
\begin{equation}
\delta_1\  \sim\ e^{-k_2(2L+1)}\ \ 
\label{scaling}
\end{equation}
where $(2L+1)$ is precisely the scattering volume length from probe I to 
probe II, and $k_2$ is the momentum corresponding to the second subband 
energy ${\cal E}_2$. We have plotted $-ln[\delta_1/(2L+1)]/k_2$ in Fig. (5b)
for several energies.  Our numerical data supports Eq. (\ref{scaling})
quite well for large $L$, and for energies closer to ${\cal E}_2$. It is not
difficult to understand the form of Eq. (\ref{scaling}).  Due to the 
scattering at the junction where the three probes meet, complicated mode
mixing takes place.  While the incoming electron is in the first subband,
mode mixing generates wavefunctions for many higher subbands, including
the second subband, which become evanescent in the probes.
For a scattering volume with a small $L$, the evanescent mode may 
``leak'' out of the volume.  However when we calculate the GPDOS from 
the scattering matrix, these ``leaked'' evanescent modes are not explicitly 
included, leading to a finite $\delta_{\beta}$.  As we increase
$L$, the evanescent modes decays away, and $\delta_{\beta}$ is reduced. 
In a specific example which can be solved exactly\cite{wang2}, a similar
form to Eq.(\ref{scaling}) was derived which was needed to correct the
GPDOS in order to satisfy the precise current conservation.  Our numerical
study presented here reinforces the results of Ref. \cite{wang2}.

To further investigate the finite size effect to GPDOS, 
in Fig. (6a,b) we plot the total DOS as obtained by GPDOS and LPDOS
as functions of energy, for three system sizes $L$.
The current conservation condition is satisfied very well for most 
of the first subband energies. When approaching the end of first subband, 
the current conservation condition is violated gradually, {\it i.e.}
$\delta_{\beta}\neq 0$. We see that for the smallest 
scattering region $L=0$, the agreement of the two total DOS 
is at best reasonable when the incident electron is from 
probe I and is away from the second subband edge (Fig. (6a)), 
and is quite bad when the electron is coming from probe III (Fig. (6b)).
The situation improves considerablely when we increased the system size.
As shown in Fig. (6), for $L=1$ and $L=2$, the agreement of the two
total DOS are much better.  However there is always a divergent behavior 
near the second subband for all sizes examined if the energy is made 
very close enough to ${\cal E}_2$.  Hence the effect of increasing the 
size of the scattering volume is to decrease the ``critical region'' 
where the two total DOS disagrees.

\section{Summary}

In conclusion, the low frequency quantum transport properties of a 
three-probe mesoscopic conductor are studied using B\"uttiker's current 
conservation formalism. The static transmission coefficients and emittance 
matrix of the system with different electric potentials are computed. 
We found that the behavior of the emittance matrix is closely
related to that of the transmission coefficients. We examined the
finite size effect of the GPDOS which affects the electric current
conservation. In general as the incoming electron energy $E$ approaching 
the threshold of the second subband, the finite-size GPDOS 
diverges and the current conservation is violated. The effect of 
increasing the size of the scattering volume is to decrease the 
region where the current conservation is violated.

\section*{Acknowledgments}

We gratefully acknowledge support by a RGC grant from the Government of Hong
Kong under grant number HKU 261/95P, a research grant from the Croucher
Foundation, the Natural Sciences and Engineering
Research Council of Canada and le Fonds pour la Formation de Chercheurs et
l'Aide \`a la Recherche de la Province du Qu\'ebec. We thank the Computer
Center of the University of Hong Kong for computational facilities and
the access of SP2 supercomputer.

\newpage
\section*{Figure Captions}

\begin{itemize}

\item[{Figure 1.}] Schematic plot of the 3-probe quantum wire system. 
The scattering volume is defined by the dotted lines.

\item[{Figure 2.}] 
The transmission coefficients and the emittance $E_{\alpha\beta}$ as
functions of the incoming electron energy without the tunneling barriers.  
(a). Solid line: reflection coefficient $R_{11}$; dotted line: 
transmission coefficient $T_{21}$; dashed line: $T_{31}$. 
(b). Emittance $E_{\alpha\beta}$. Solid line: $E_{11}$; dotted line:
$E_{21}$; dashed line: $E_{31}$.

\item[{Figure 3.}] 
Electron dwell time $\tau_1$ as a function of the incoming electron 
energy in the tunneling regime. The three peaks indicate three resonance
states in the system in this energy range.  Inset: $\tau_3$.

\item[{Figure 4.}] 
The transmission coefficients and the emittance $E_{\alpha\beta}$ as
functions of the incoming electron energy in the tunneling regime.
(a). Solid line: $R_{11}$; dotted line: $E_{11}$; 
(b). Solid line: $T_{21}$; dotted line: $E_{21}$;
(c). Solid line: $T_{31}$; dotted line: $E_{31}$.

\item[{Figure 5.}] 
(a).  The difference, $\delta_1$, of the total PDOS as computed from the 
GPDOS and LPDOS from Eq. (\ref{delta}) as a function of the scattering
volume linear size $L$.  Solid line: at energy $E=39.46699$; dotted line:
at energy $E=39.45371$; dashed line: at energy $E=39.44137$;
(b). The quantity, $-ln[\delta_1/(2L+1)]/k_2$ as a function of the linear
size $L$ for several incoming electron energies as shown.  At large $L$, 
this quantity approaches unity, confirming the form of Eq. (\ref{scaling}).

\item[{Figure 6.}] 
Comparison of the total PDOS computed from the GPDOS and LPDOS, as a
function of the incoming electron energy for three different sizes $L=0,
1$, and $2$ in the transmissive regime.
(a). Electrons coming from probe 1.  (b). Electrons coming from probe 3.
The agreement of the total PDOS is quite good up to the ``critical region''
near the onset of the second subband.

\end{itemize}

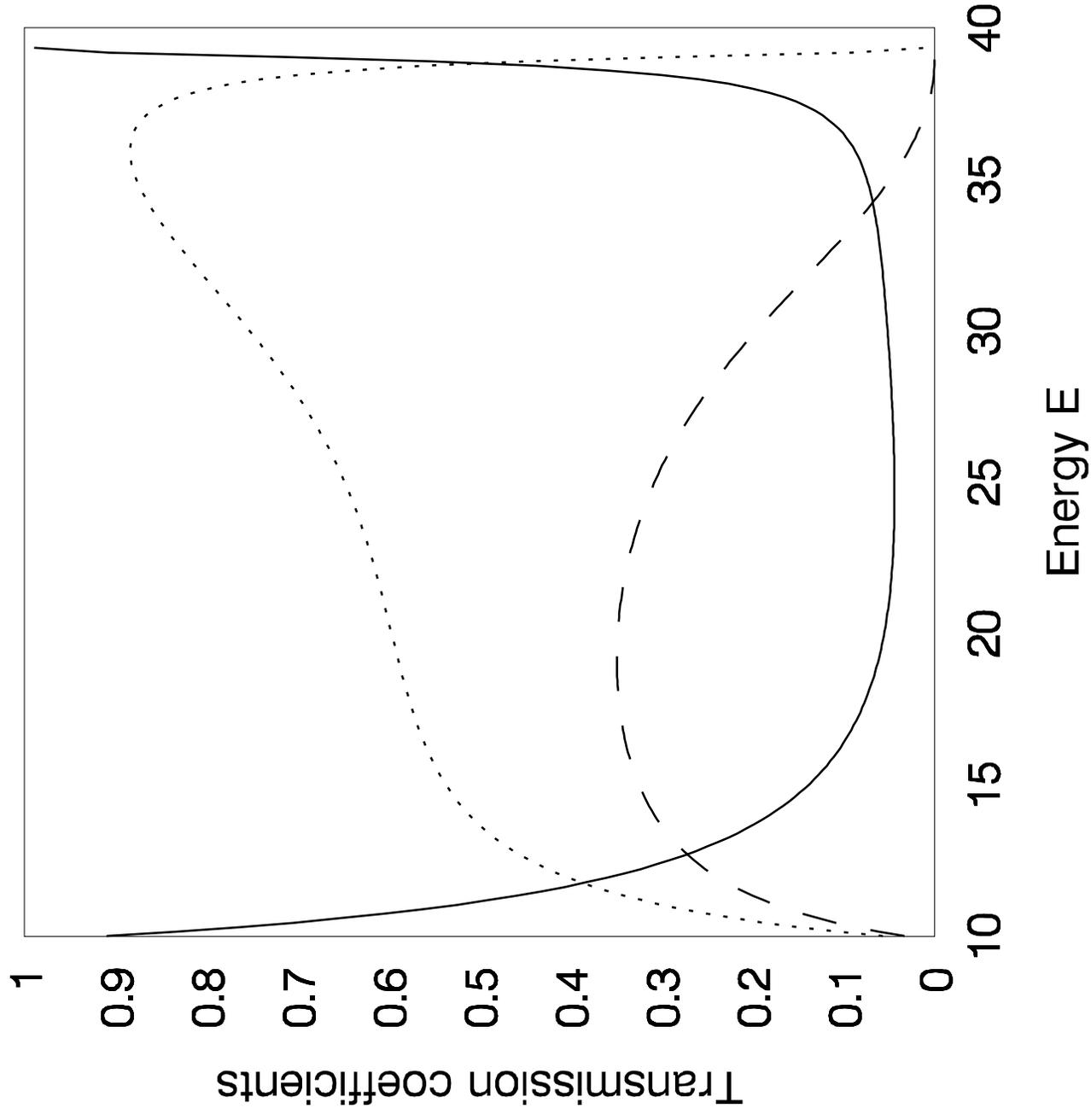
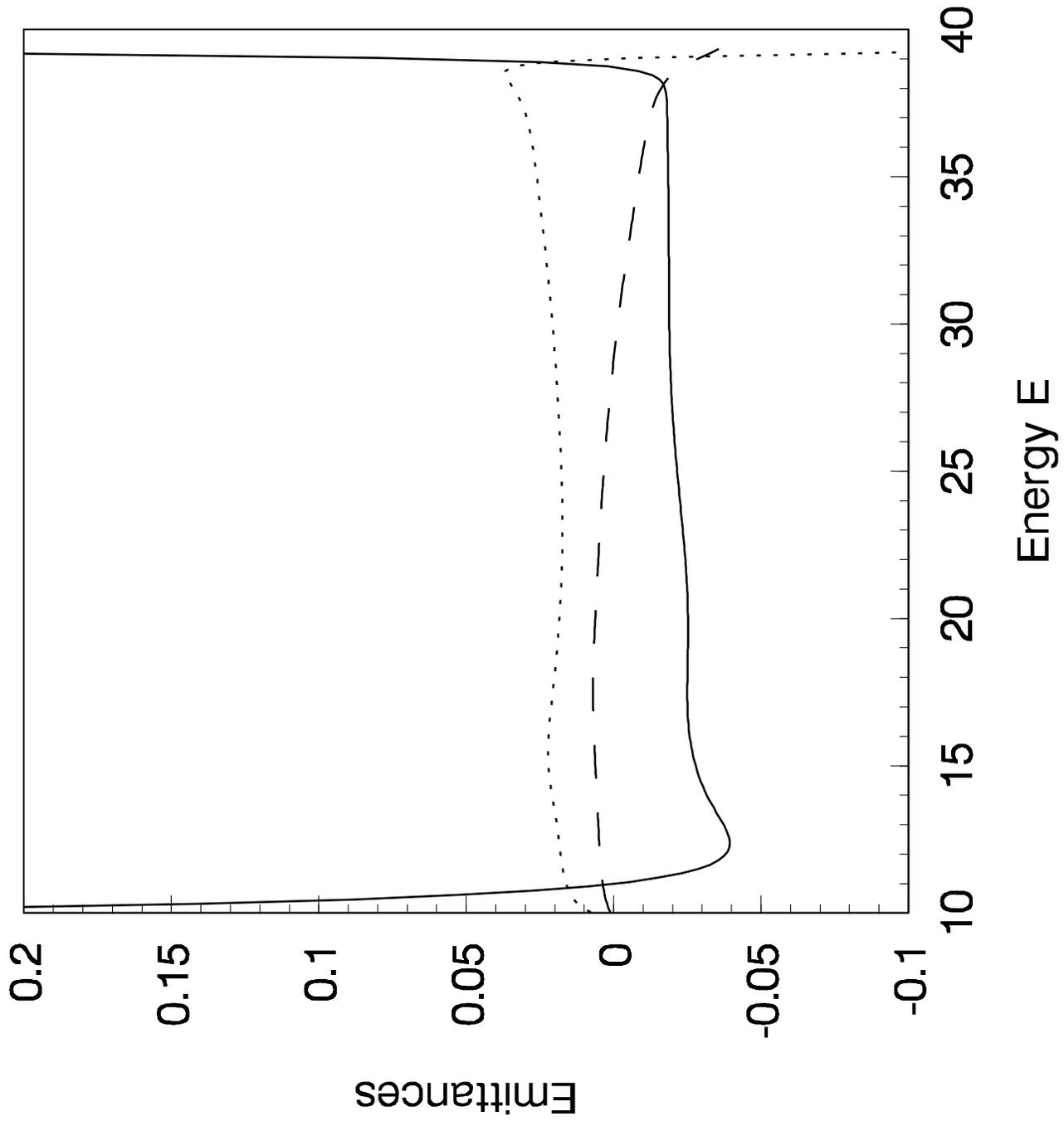
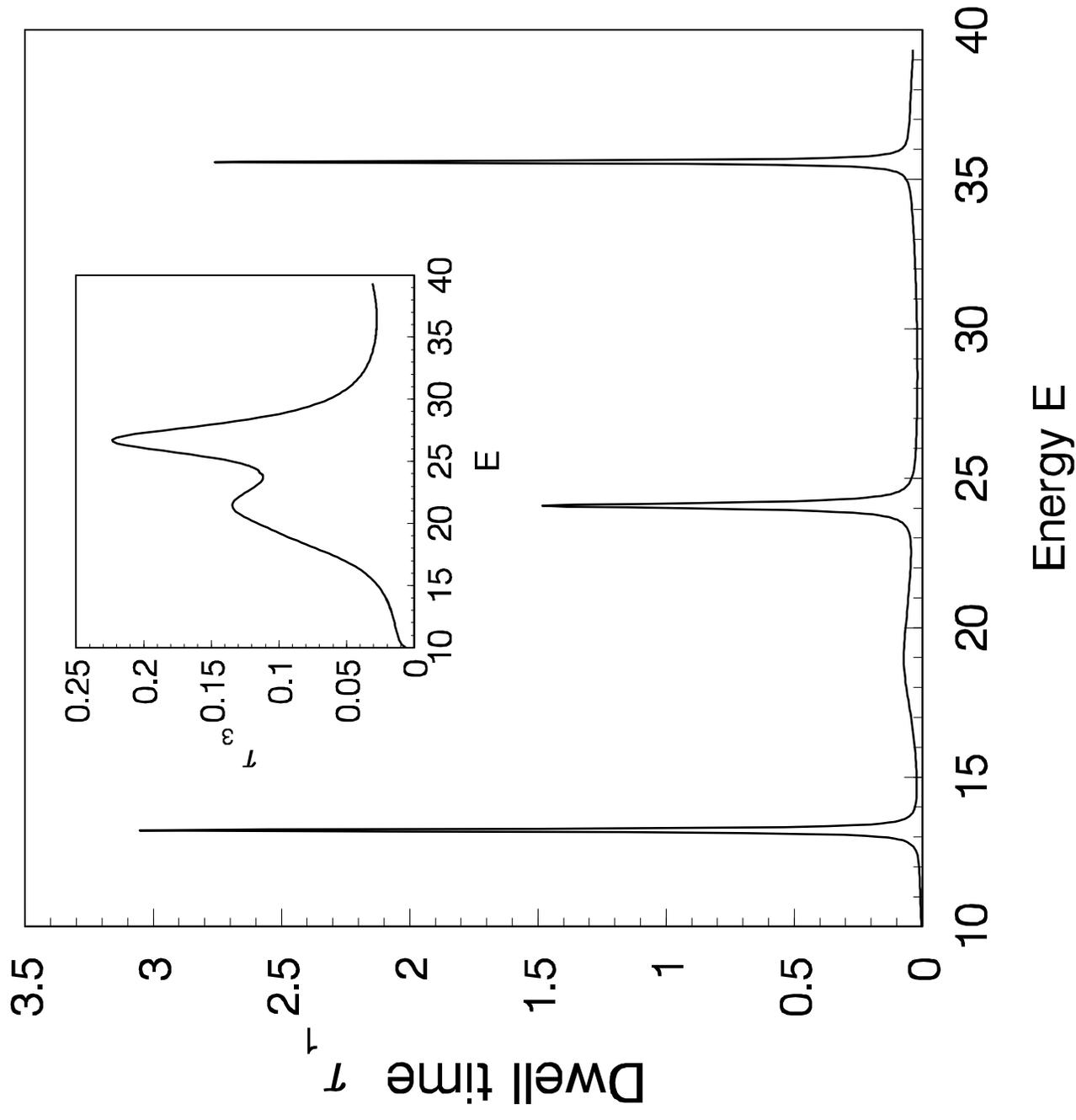
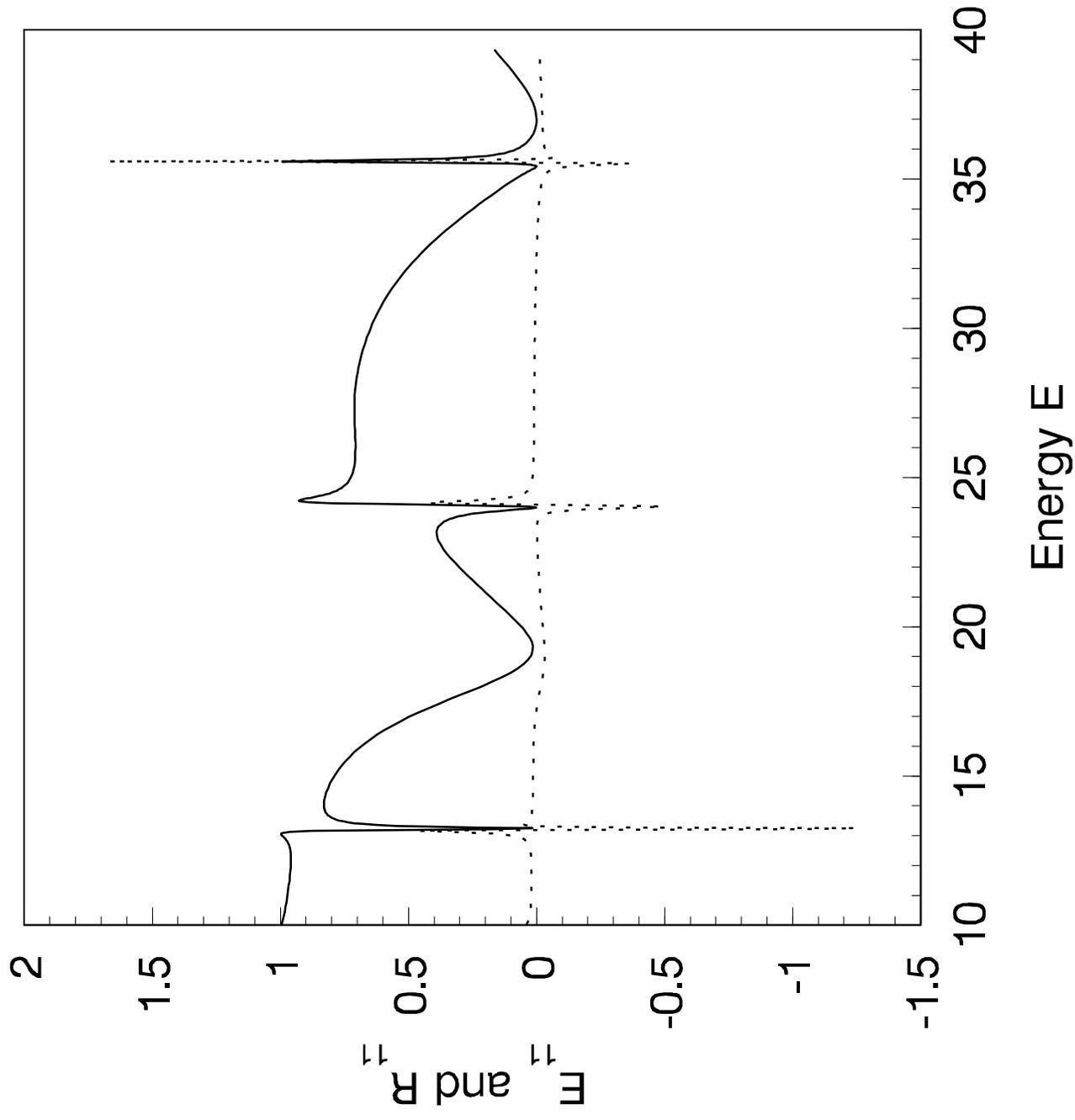
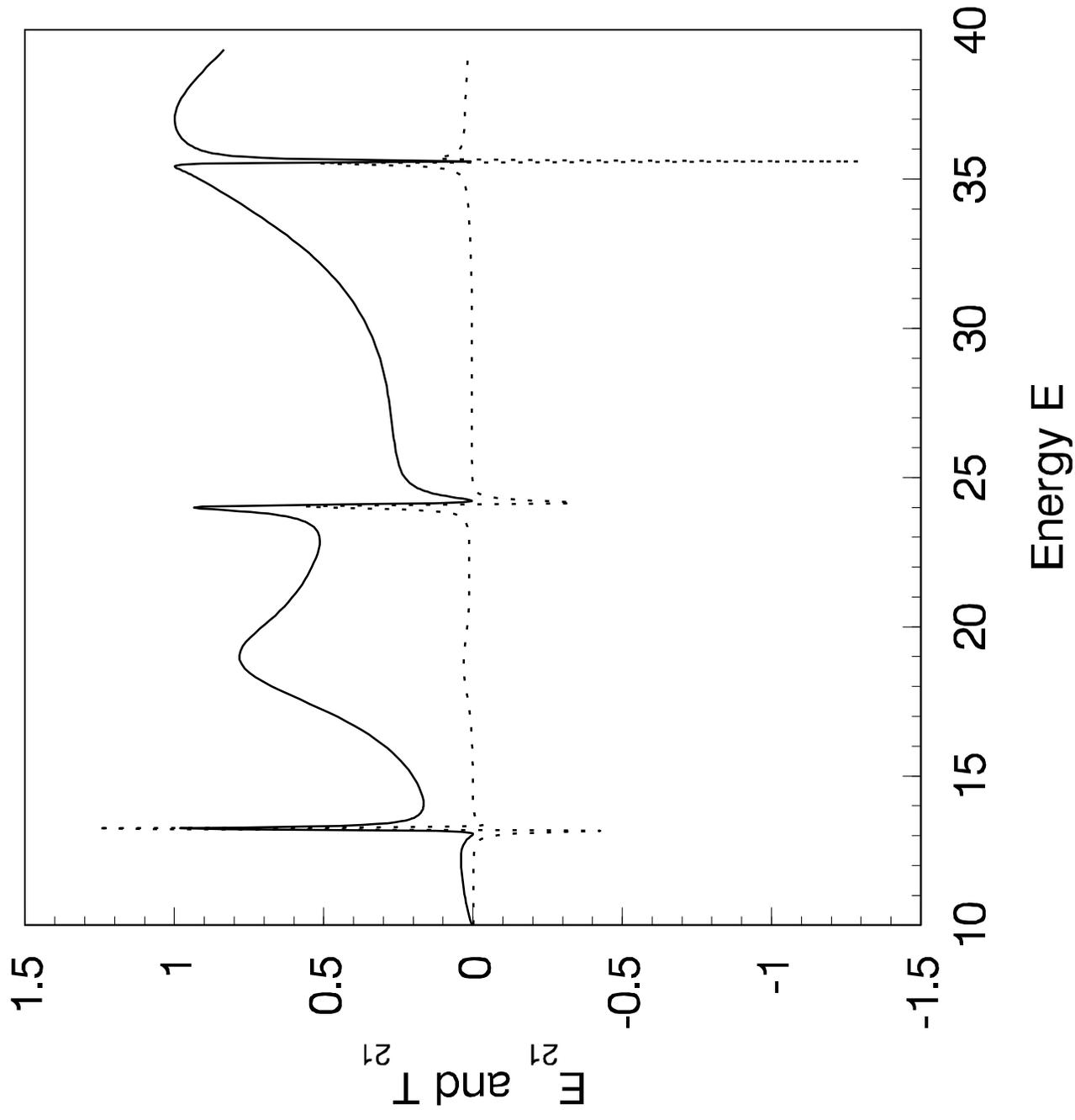
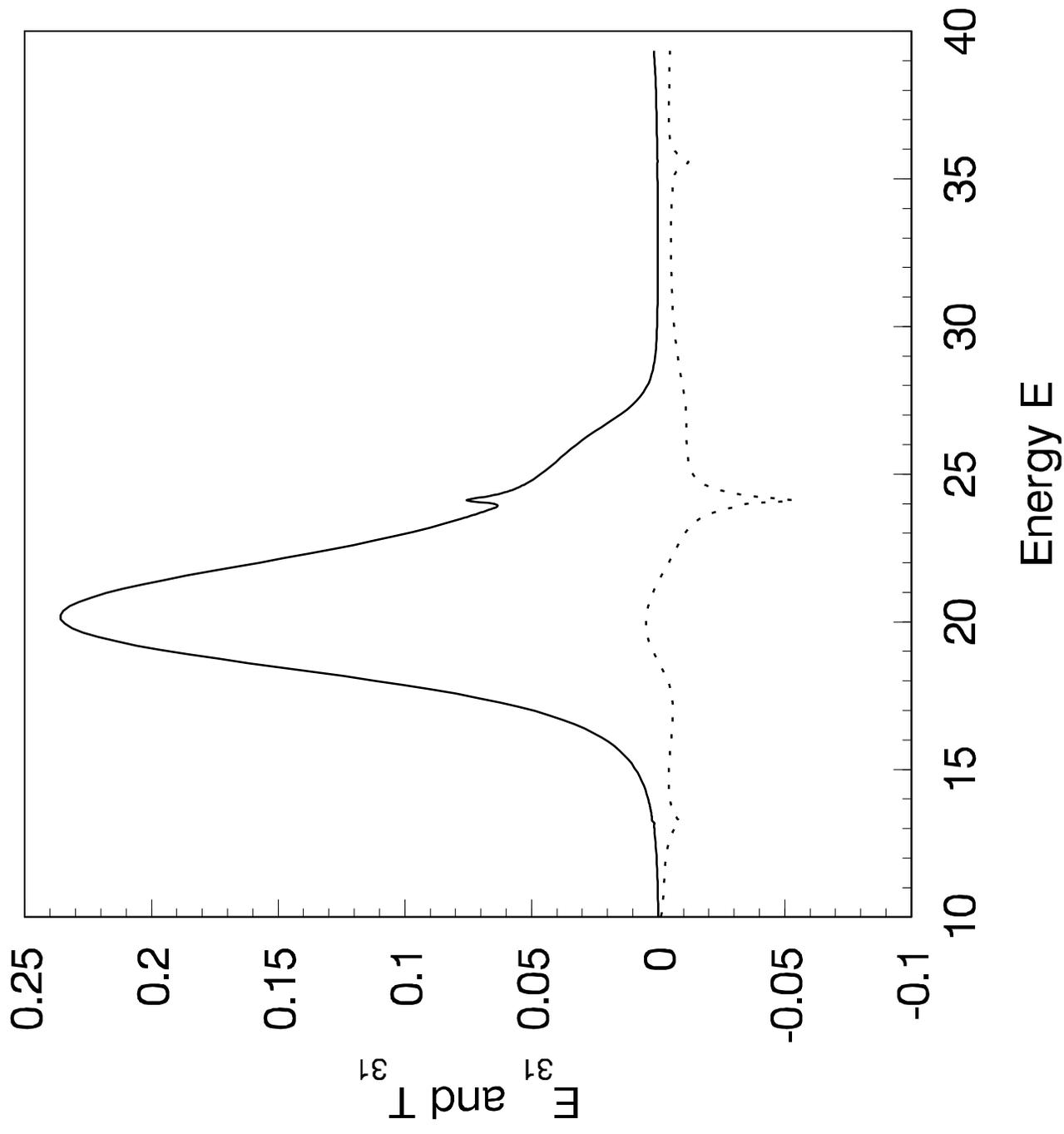
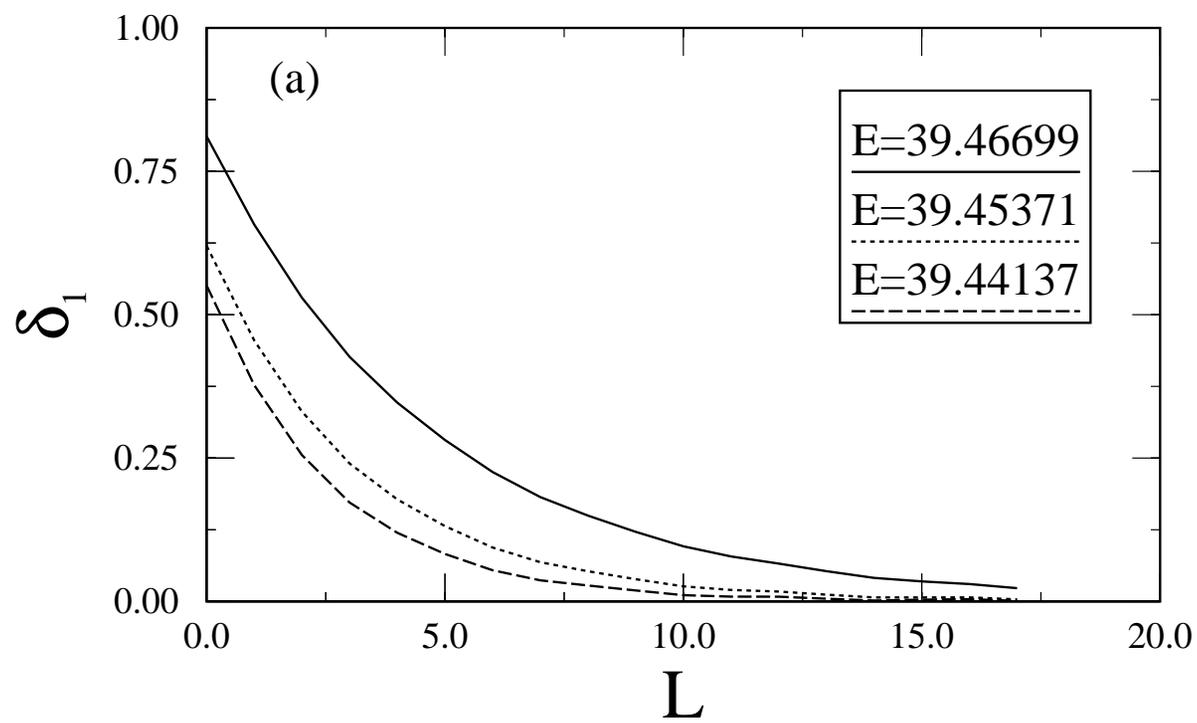
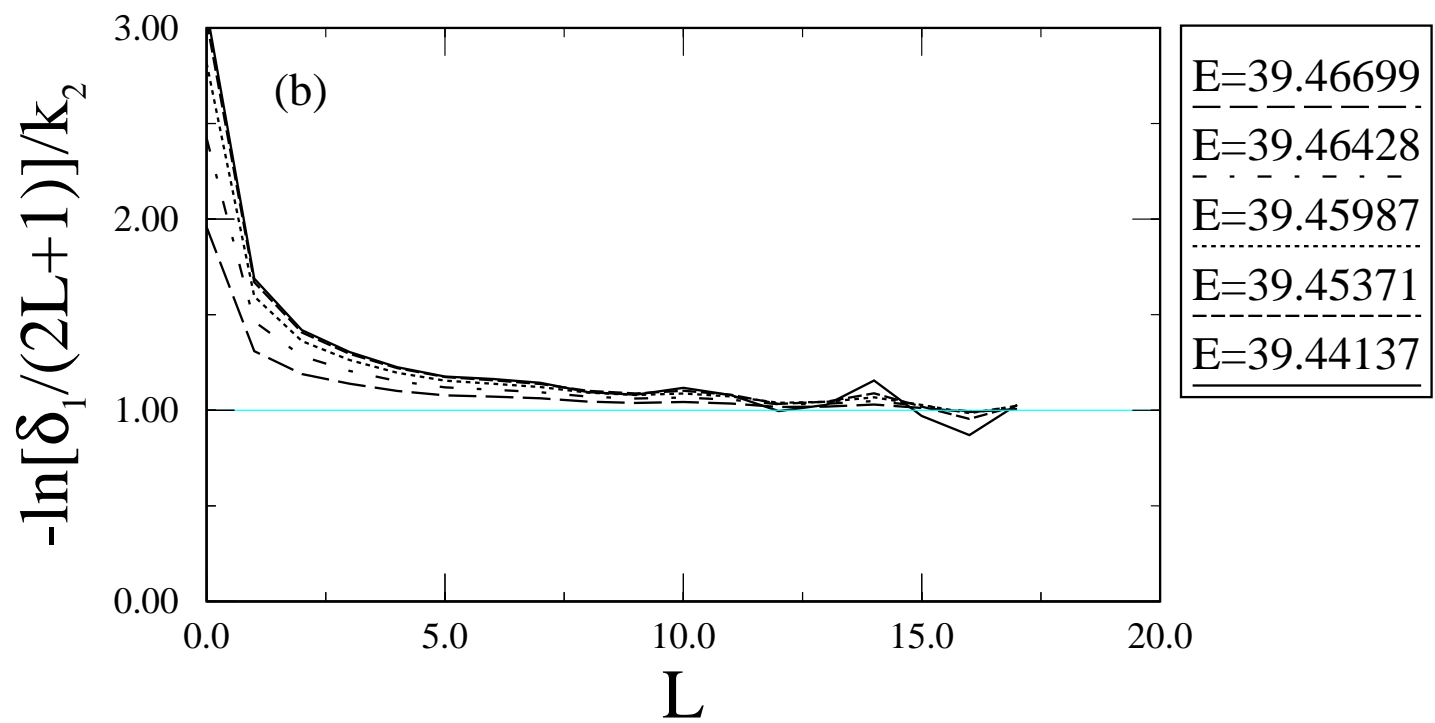
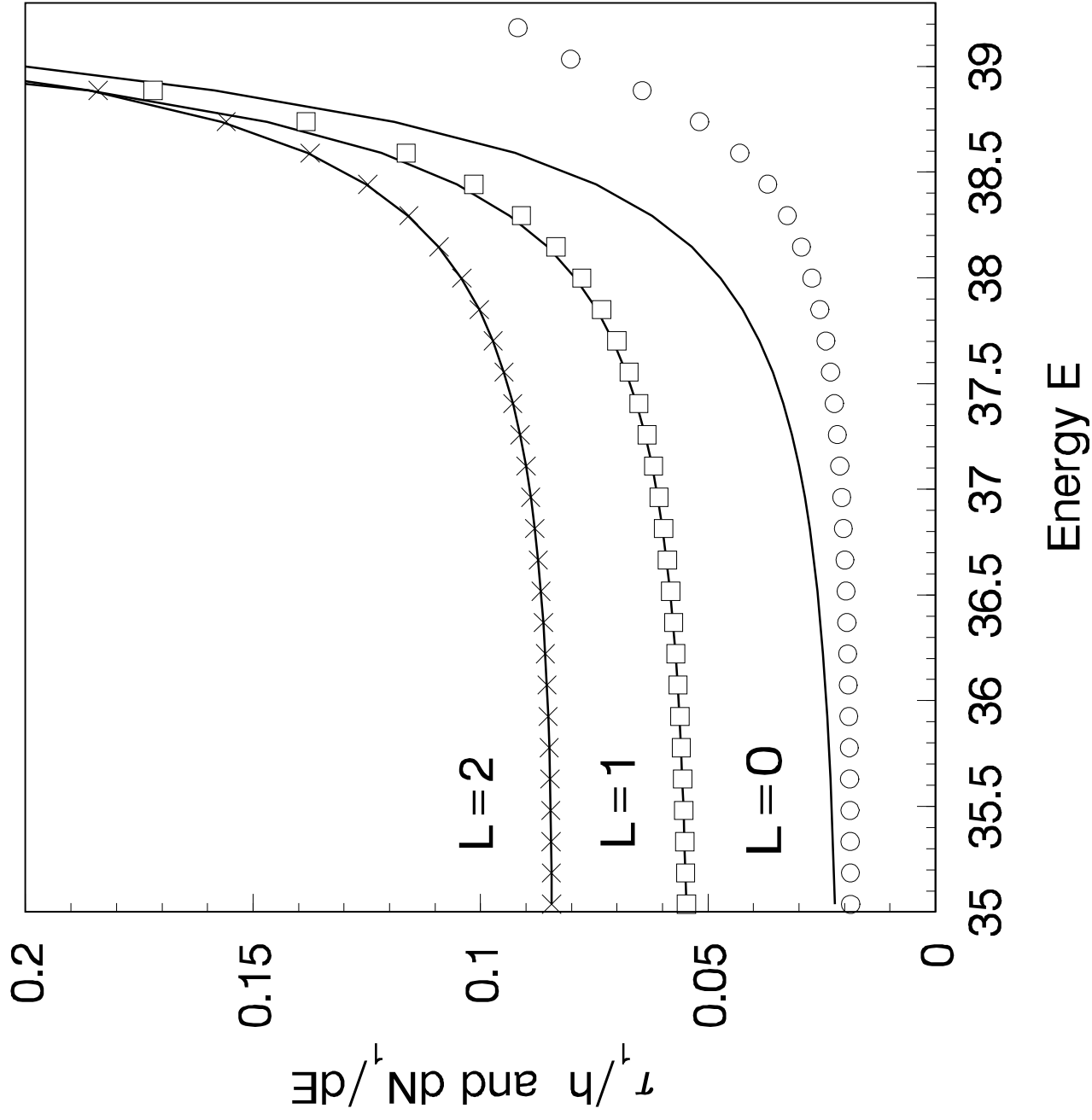
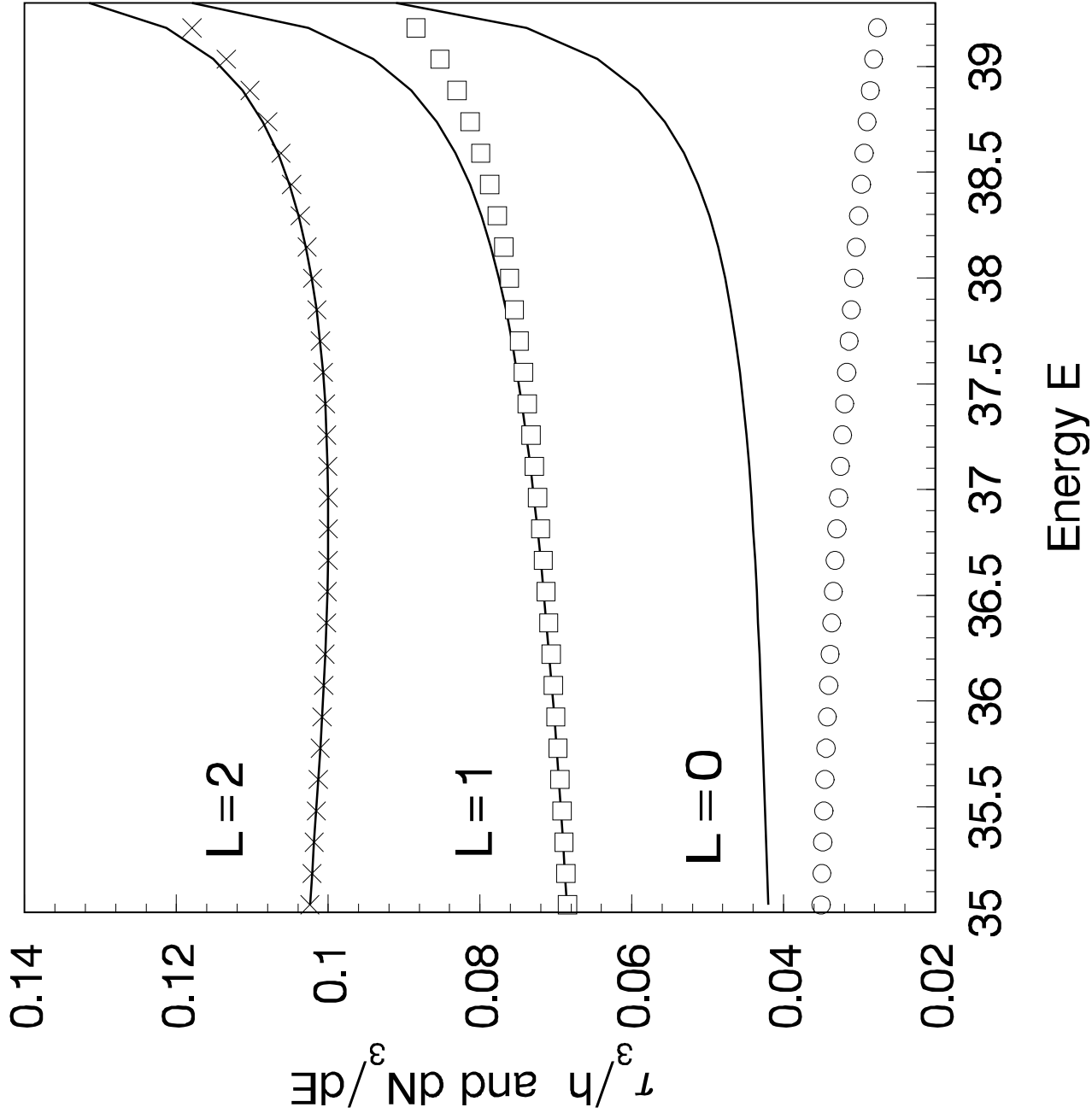
\end{document}